\documentclass{iaus}
\usepackage{graphicx}
\usepackage{natbib}

\title[Massive star nucleosynthesis] 
{Nucleosynthesis in Rotating massive stars and Abundances in the Early Galaxy}

\author[Meynet et al.]   
{Georges Meynet$^1$,
Raphael Hirschi$^{2,3}$,
Sylvia Ekstrom$^1$,
Andr\'e Maeder$^1$,
Cyril Georgy$^1$,
Patrick Eggenberger$^1$,
Cristina Chiappini$^1$}

\affiliation{$^1$Geneva Observatory, Geneva University, \\ CH--1290 Sauverny, Switzerland \\ email: {\tt georges.meynet@unige.ch} \\[\affilskip]
$^2$Astrophysics group, Keele University, \\ Lennard-Jones Lab., Keele, ST5 5BG, UK 
\\ email: {\tt r.hirschi@epsam.keele.ac.uk }\\[\affilskip]
$^3$IPMU, University of Tokyo, \\ Kashiwa, Chiba 277-8582, Japan}

\pubyear{2009}
\volume{265}  
\pagerange{119--126}
\jname{Chemical Abundances in the Universe: Connecting First Stars to Planets}
\editors{K. Cunha, M. Spite \& B. Barbuy, eds.}
\begin{document}

\maketitle

\begin{abstract}
We discuss three effects of axial rotation at low metallicity. The first one is the mixing of the chemical species which is predicted
to be more efficient in low metallicity environments. A consequence is the production of important quantities of primary $^{14}$N,
$^{13}$C, $^{22}$Ne and a strong impact on the  nucleosynthesis of the  {\it s}-process elements. The second effect is a consequence of the first.
Strong mixing makes possible the apparition at the surface of important quantities of CNO elements. This increases the opacity of the
outer layers and may trigger important mass loss by line driven winds. The third effect is the fact that, during the main-sequence phase, 
stars, at very low metallicity, reach more easily than their more metal rich counterparts, the critical velocity\footnote{The critical velocity
is the surface equatorial velocity such that the centrifugal acceleration compensates for the local gravity.}. We discuss the respective importance
of these three effects as a function of the metallicity. We show the consequences for the early chemical evolution of the galactic halo and
for explaining the CEMP stars. We conclude that rotation is probably a key feature which contributes in an important way to shape
 the evolution of the first stellar generations in the Universe.
\end{abstract}

\firstsection 
\section{Introduction}

The study of the most iron poor stars in the halo offers a unique window on the nature and the evolution of the
first generations of stars which provided the matter from which these halo stars were formed.
Constraints on the first stellar generations may provide interesting views on topical questions as the nature
of the stars which reionize the Universe, the initial mass function of the first stellar generations, the frequency of long soft Gamma Ray Bursts at very low metallicity,
the timescale for the mixing of the newly synthesized products with interstellar medium material,  or the
conditions of star formation in the very early life of our Galaxy. 

\begin{figure}[t]
\includegraphics[width=2.4in,height=2.4in,angle=0]{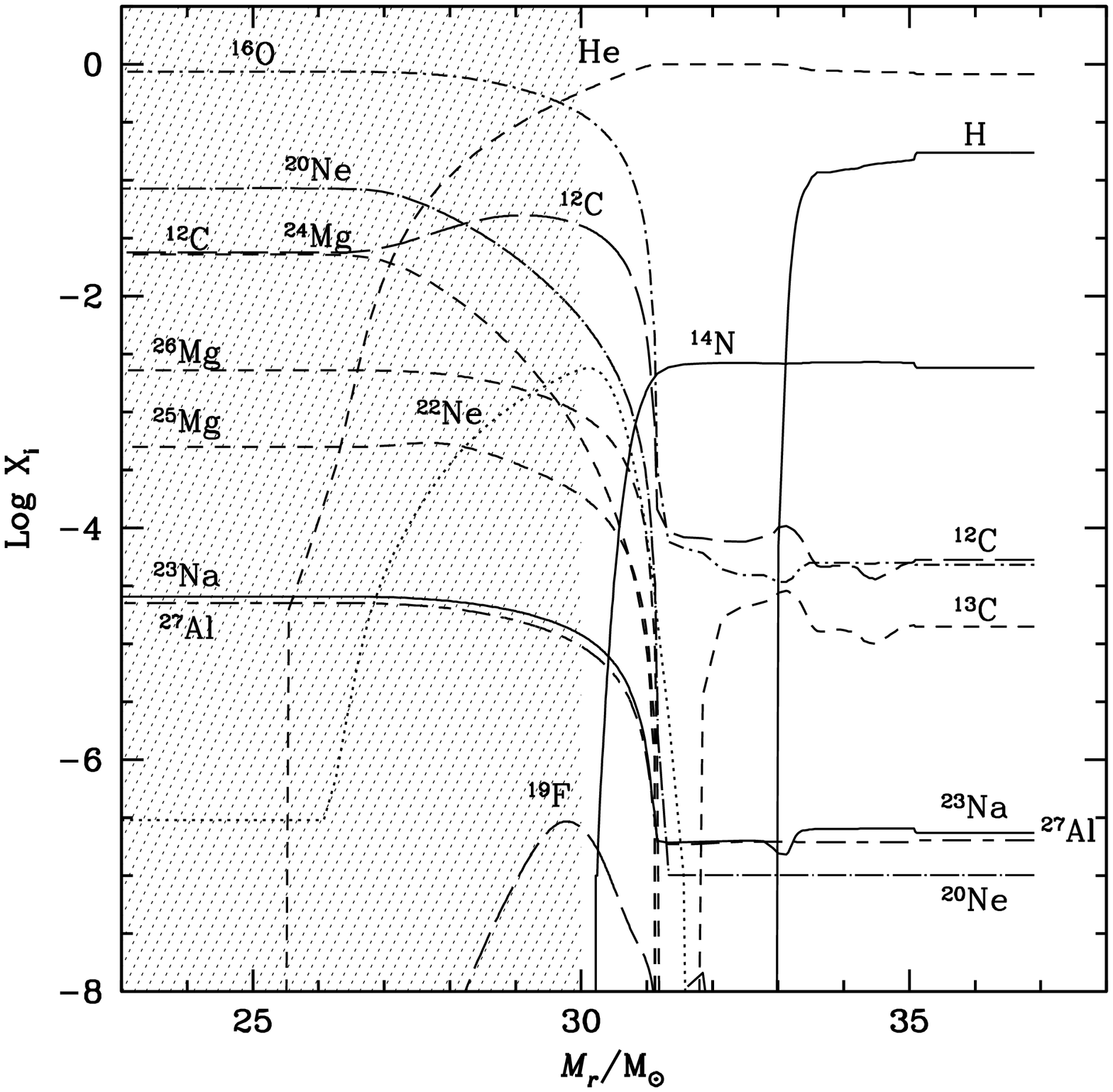}
\hfill
\includegraphics[width=2.4in,height=2.4in,angle=0]{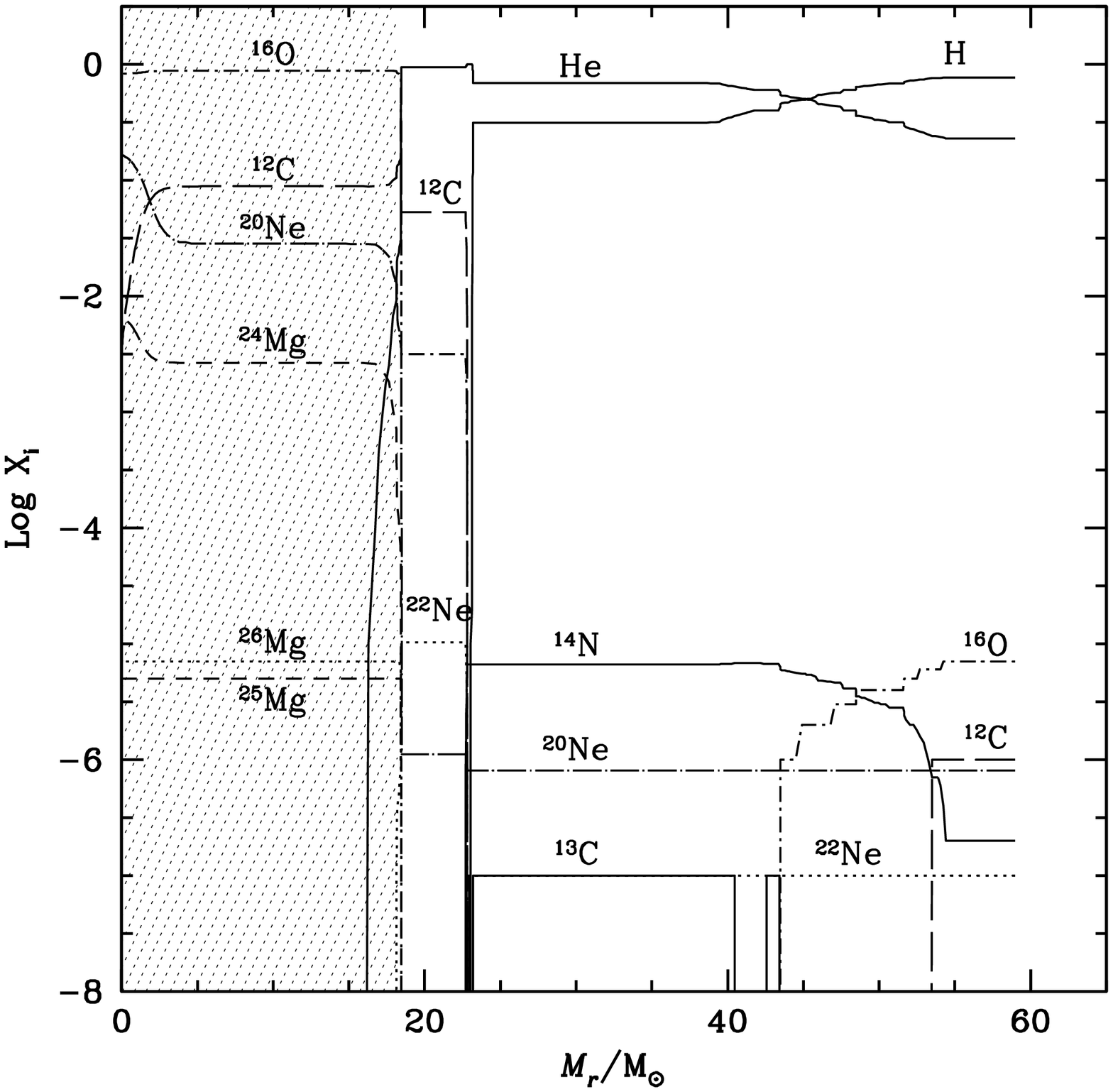}
\caption{{\it Left panel} : variation of the abundances of various elements  (in mass fraction) as a function of the Lagrangian mass in the outer layers
     of a 60 M$_\odot$ model at $Z=10^{-5}$ with $\upsilon_{\rm ini}=800$ km s$^{-1}$ at the end of the core He-burning phase. The grey area 
     covers the CO core. {\it Right panel :} same as left part for a non rotating 60 M$_\odot$ model at $Z=10^{-5}$. Models computed by \citet{paperVIII2002} and \citet{Meynet2006}.}
\label{struc}
\end{figure}

In this context, many observed features of the very iron poor stars are quite surprising and were not at all
expected:
\begin{itemize}
\item One expected that halo stars, having a very low [Fe/H], should show an important scatter in their abundances. Indeed, at these very early times, stars are expected to form from not well
mixed clumps and thus should bear the nucleosynthetic signatures of a few peculiar events. But
 \citet{Cayrel2004} find that the scatter in the abundances of several elements ratios (e.g. [Cr/Fe]) is very small. It can be as low as 0.05 dex.
In contrast, r-process elements and nitrogen for instance reveal large star-to-star scatters \citep{Ryan1996, Honda2004, Andrievsky2009}. 
Explanation are proposed in  \citet{Ishimaru2004, Ishimaru2006, Cescutti2008}.  
\item One expected that the very first generations of Pop. III stars were very massive and contributed to the enrichment
of the interstellar medium through Pair Instability Supernovae \citep[PISN,][]{Barkat1967, Bond1984}. The nucleosynthetic pattern produced by such events is well understood and presents specific features  \citep{HegerWoosley2005}, which were expected to be observed at least in some iron-poor halo stars. At the moment no trace of PISN has been found  \citep{Cayrel2004}.  Some authors have suggested an explanation why this signature has, up to now, escaped detection  \citep{Karlsson2008, Ekstrom2008d}.
\item Spectroscopic observations of halo stars
 \citep[e.g.][]{Spite2005} indicate a primary production
 of nitrogen over a large metallicity range at low metallicity.  According to the chemical evolution models of \citet{Chiappini2006, Chiappini2008},
 primary nitrogen production by non-rotating or slow-rotating Pop III stars is not sufficient to explain the observations. For instance, primary nitrogen needs to be produced
 on larger ranges of masses and metallicities than expected in non rotating standard models.
 A similar plateau is observed in Damped Lyman Alpha (DLA) systems \citep{Pettini2008}. 
 \item  Halo stars with log(O/H)+12 inferior to about 6.5 present higher C/O ratios than
 halo stars with log(O/H)+12 between 6.5 and 8.2 \citep{Akerman2004, Spite2005}.
Again a similar trend is observed in DLAs  \citep{Pettini2008}. This is not predicted by slow rotating models.
\item The features listed above concern the bulk of the halo field stars. Now in addition to this ``normal'' population, there exists a group of stars showing a very different composition. These stars are collectively called Carbon-Enhanced Metal Poor stars (CEMP) and present high [C/Fe] ratios  \citep[see the review by][]{Beers2005}.  The scatter in [C/Fe] is quite important indicating that these stars were not formed from a well mixed reservoir but acquired their peculiar high carbon abundance from
locally enriched material. Some of these stars show also strong overabundances (with respect to iron) of nitrogen, oxygen and other
heavy elements.  
At first order, these stars do appear to be formed from material having been processed mainly by H- and He-burning processes.
\item Very interestingly, halo stars in clusters also present some surprising features. While part of the cluster population follows the same trend as the ``normal'' population of the field, another part presents very different chemical patterns, with
for instance, strong depletions of oxygen and strong enhancements in sodium \citep[see the review by][]{Gratton2004}. These stars do appear to be formed
from material having been processed only by H burning.
\end{itemize}
Valid models should provide explanations for all these features. Ideally they should also provide some predictions for not yet observed characteristics. In the following
we discuss the possible role of stellar axial rotation and we show how it can deeply affect the nucleosynthetic outputs of
stars both in the massive ($>$ 8 M$_\odot$) and intermediate mass range (2 $<$ M/M$_\odot$ $<$ 8) and how it can explain some of the above observed features.

\section{Effects induced by rotation}

\begin{table}
  \begin{center}
  {\scriptsize
  \caption{Comparison of the outputs of various stellar models.}
  \label{tab1}
 {\scriptsize
  \begin{tabular}{ccccccl}
  \hline 
 Mass              & $Z$                               & $\upsilon_{\rm ini}$  & $\Delta M_{\rm MS}$ &   $\Delta M_{\rm post-MS}$                 &  Max. of X$_{\rm s}(^{14}$N) & Reference \\ 
 M$_\odot$    &                                        & km s$^{-1}$                &          M$_\odot$          &        M$_\odot$                                       &                                                        &                      \\  
\hline
60                   & 0                                     & 0                                   &  0.00                              & 0.00                                                           &  0.00                                              & \citet{Ekstrom2008a}                       \\
60                   & 10$^{-8}$                      & 0                                   &  0.18                              & 0.09                                                          &  2.34 10$^{-10}$                                              & \citet{Meynet2006}                       \\
60                   & 10$^{-5}$                      & 0                                   &  0.21                              & 0.22                                                           &   2.34 10$^{-7}$                                             & \citet{Meynet2006}                       \\
60                   & 5 $\times$ 10$^{-4}$  & 0                                   &  0.78                              & 13.29                                                         &  2.21 10$^{-4}$                                              & \citet{Decressin2007a}                       \\
                       &                                         &                                      &                                        &                                                                     &                                                        &                                                           \\
60                   & 0                                     & 800                               &  2.32                              & 2.41                                                           &  3.54 10$^{-4}$                                             & \citet{Ekstrom2008a}                       \\
60                   & 10$^{-8}$                      & 800                               &  2.38                              & 33.64                                                         &  1.02 10$^{-2}$                                              & \citet{Meynet2006}                       \\
60                   & 10$^{-5}$                      & 800                               &  6.15                              & 16.57                                                         &  2.07 10$^{-3}$                                             & \citet{Meynet2006}                       \\
60                   & 5 $\times$ 10$^{-4}$   &800                               &  20.96                             & 21.79                                                         & 3.86 10$^{-2}$                                              & \citet{Decressin2007a}                       \\
 \hline
  \end{tabular}
  }
  }
 \end{center}
 \end{table}

The physics of rotation included in present stellar models has been recently exposed in \citet{Maeder2009}. Here we just
briefly recall the main effects. First rotation triggers many instabilities in stellar interiors. These instabilities  participate in the transport of chemical species and of angular momentum in, otherwise, stable, radiative regions.

Around solar $Z$, rotating models improve a lot the agreements between the predictions of the stellar models and the observations. For instance the changes of the surface chemical abundances can be reproduced when the effects of rotation are accounted for \citep{Maeder2009a}. Also the variation with the metallicity of the number ratio of Wolf-rayet to O-type stars
can be reproduced \citep{paperXI2005}. When the same physics, allowing these successes at near solar metallicity, is implemented in stellar models at very low metallicity, one notes interesting consequences. The most important consequence (the other
consequences can  actually  be deduced from this one) is that 
the stars with the lowest metallicity (other characteristics being kept the same) will be more mixed by rotational mixing  than the stars with a higher metal content \citep{paperVII2001}. 
This comes mainly from the fact that at low $Z$, stars are more compact. This makes the gradient of the angular velocity inside the star steeper
and thus the transport of the chemical species by shear mixing more efficient. From this characteristic, occurring at 
low $Z$, two consequences of great interest for the questions we are discussing here have been found: first rotating massive and intermediate mass stars
can produce important amounts of primary nitrogen. The level of production depends on the initial velocity of the models.  Other elements have their production boosted by rotational mixing, among them  are $^{13}$C, $^{22}$Ne\footnote{Enhancement in the abundance of $^{22}$Ne in the He-core of the rotating model has a strong impact on the {\it s-}process according to the
work by \citet{Pignatari2008}.}. This is shown
in Fig. \ref{struc}, where the compositions of a non-rotating and a rotating 60 M$_\odot$ model at $Z$=10$^{-5}$ are plotted. 

 A second consequence is that, during the core He-burning phase, the surface of the star can be strongly enriched in CNO elements, increasing its surface metallicity by many orders of magnitudes
 \citep{paperVIII2002, Hirschi2007}. This
strong surface metallicity enhancement can trigger important mass losses driven by  radiatively line driven winds.
The material ejected in that way will bear the nucleosynthetic signature of both H- and He-burning processes.

Another effect which occurs preferentially at low $Z$ is the loss of mass through mechanical mass loss.
What we call here mechanical mass loss is the loss of mass when the surface velocity is so high that,
at the equator, the centrifugal acceleration is equal to the gravity. When such circumstances are realized, a tiny kick suffices to launch matter into a keplerian orbit around the star and thus to form an equatorial disk. What does happen
to that material remains quite speculative. A reasonable hypothesis is that this material is lost. 
This process does occur in near solar metallicity for sufficiently rapidly rotating B-type  stars. Not for O-type stars, because, at solar metallicity, they lose too much mass and therefore angular momentum by stellar winds \citep{Ekstrom2008b}.
At very low metallicity, one expects that the reaching of the critical limit will also occur for more massive stars as  O-type stars, because, in contrast to what happens at high metallicities, radiatively driven winds are weaker during the
MS phase and thus do not remove the angular momentum brought to the surface by the meridional currents.

What is the importance of these various effects? How do they vary as a function of the initial mass, metallicity
and initial rotational velocities? Some elements of response can be obtained from Table \ref{tab1}. We can note the following trends
\begin{itemize}
\item The mass lost during the MS phase ($\Delta M_{\rm MS}$), in rotating models, is due, at these very low metallicities, mainly to the reaching
of the critical limit (see column 4 of Table \ref{tab1}).  The effect of mechanical mass loss can be estimated by comparing
$\Delta M_{\rm MS}$ for the rotating and the non-rotating models.
We see that the mechanical mass loss has  a
kind of metallicity dependence. The model at higher metallicity loses much more mass by this process
than models at lower metallicities. The reason for this is due to three facts: first, a given value of the velocity on the ZAMS
corresponds, at high $Z$, to a higher value of the ratio $\upsilon/\upsilon_{\rm crit}$  than at low $Z$ \citep[see Fig. 12 in][]{Ekstrom2008b}.
Second, the meridional currents slow down the
internal regions and accelerate the surface and are thus the main agents which bring the velocity of the surface
near the critical one. Third, these currents are slower in more compact stars. Therefore, in low metallicity stars,
it takes more time, starting from a given initial
velocity, to reach the critical limit \footnote{Let us note that, due to these three effects, without mass loss, stars would reach the critical limit
more rapidly at high metallicity. However,
when the metallicity is near the solar metallicity, 
the strong line driven stellar winds remove
rapidly the angular momentum brought to the surface by the meridional currents and prevent the high mass losing
stars to reach the critical limit.}.
When the critical limit is reached, the surface velocity remains near the critical value until the end of the MS phase.
This is due to the following behaviour:  the star encounters the critical limit for the first time, mass is lost at the equator, angular momentum is removed,
the star evolves away from the critical limit. Then meridional currents bring again angular momentum to the surface making
the star to evolve back against the critical limit. The cycle begins again. The timescale for the evolution back to the critical limit
will be shorter at higher than at lower metallicities for the same reason as above, and thus this will favor higher mass losses by this process.
The material released by this mechanical mass loss is only enriched in H-burning products, since
it is ejected during the MS phase.
\citet{Decressin2007a, Decressin2007b} have studied the consequences of such models for explaining the 
chemically peculiar stars observed in globular clusters.
\item In the models at $Z$ equal to 10$^{-8}$ and 10$^{-5}$, a phase exists where the star is in the red part of the HR diagram,
and presents strong enhancements of its surface CNO content. The last column of Table \ref{tab1} indicates the maximum value
of the abundance of $^{14}$N reached at the surface. In non-rotating models, no change of the surface abundance occurs for $Z \le 10^{-5}$.
In the non-rotating  $Z = 5 \times 10^{-4}$ model, nitrogen enhancement occurs as a result of mass loss. Only secondary nitrogen is produced in non metal-free, non-rotating models.
In the rotating models, primary nitrogen is produced. The most efficient producers are models with $Z=10^{-8}$ and $10^{-5}$. In these models,
the apparition at the surface of primary CNO elements
triggers high mass loss rates (see column 5 in Table \ref{tab1}). The material ejected in that
way has been processed by both H- and He-burning processes \citep{Meynet2006, Hirschi2007}. For the velocities and mass considered here, this only occurs
for non-zero very low metallicity, but not for Pop III stars.  At near solar metallicities, this does not occur
for models with the same initial angular momentum content. Why? 
For $Z$=0, we shall see below that the mixing during the core He-burning phase is
less efficient than in non-zero metallicity stellar models. At higher metallicities, rotational mixing is less efficient.
Furthermore the line driven winds become more important, and in very massive stars,  they may remove the H-burning regions 
early during the core He-burning phase preventing primary nitrogen production.
In the following, we argue that
the material ejected under the form of this wind triggered by the self enrichment of the surface, presents strong similarities
with the abundance pattern observed at the surface of the CEMP stars.
\item The rotating models produce primary nitrogen (and also primary $^{13}$C and $^{22}$Ne). The most
efficient producers are the models with $Z$ equal to 10$^{-8}$ and 10$^{-5}$. The pop III stars produce less
primary nitrogen for the following reason: due to the absence of CNO elements, these models begin to convert
H into He through the pp chains. The energy output from these chains is not sufficient to compensate for the high
luminosity of the star. Therefore the rest of the energy has to be extracted from the gravitational reservoir and the star
must contract. Contraction occurs until  the central temperatures reach sufficient high values to activate the triple alpha reaction.
Some carbon is then produced and the CNO cycle can be activated.  From this stage on, the H-burning is pursued as in more metal rich stars,
through the CNO cycle. A consequence of this is that the core H-burning occurs at temperatures very similar to that
of the core He-burning. Thus, in Pop III stars, at the end of the core H-burning phase, the core does not need to contract
a lot in order to re-activate the He-burning reactions. This absence of strong contraction maintains the star in the blue part
of the HR diagram
during most of the core He-burning phase. It also prevents a strong gradient of angular velocity to form at the border of the core
and therefore efficient shear mixing. Less primary nitrogen
is formed. In more metal rich models, less primary nitrogen is produced both because mixing is less efficient and because the H-burning
shell is more distant from the He-burning core. 
\end{itemize}

\section{Nucleosynthesis from  ``spinstars''}

\begin{figure}[t]
\includegraphics[width=2.3in,height=2.3in,angle=0]{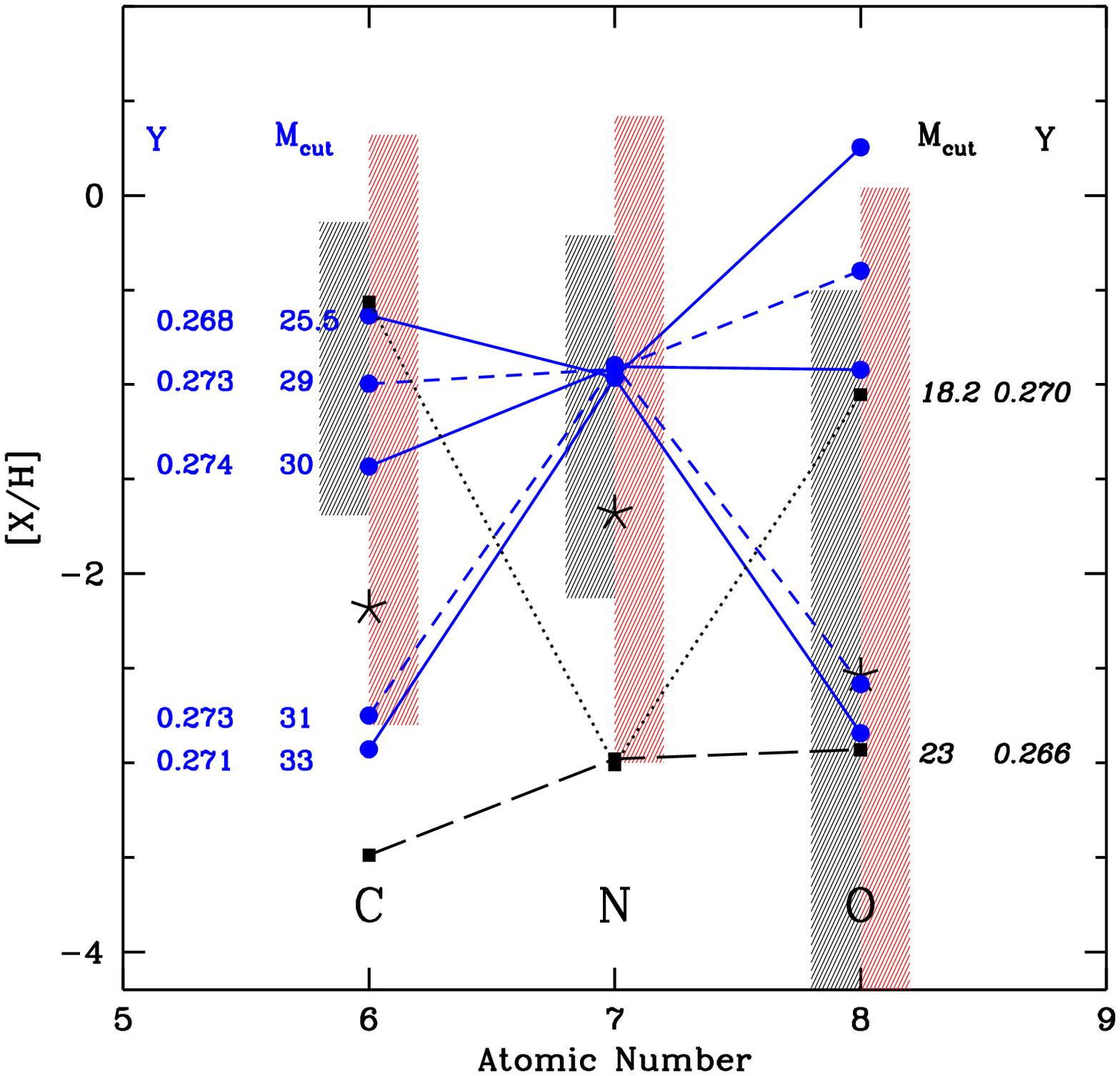}
\hfill
\includegraphics[width=2.3in,height=2.3in,angle=0]{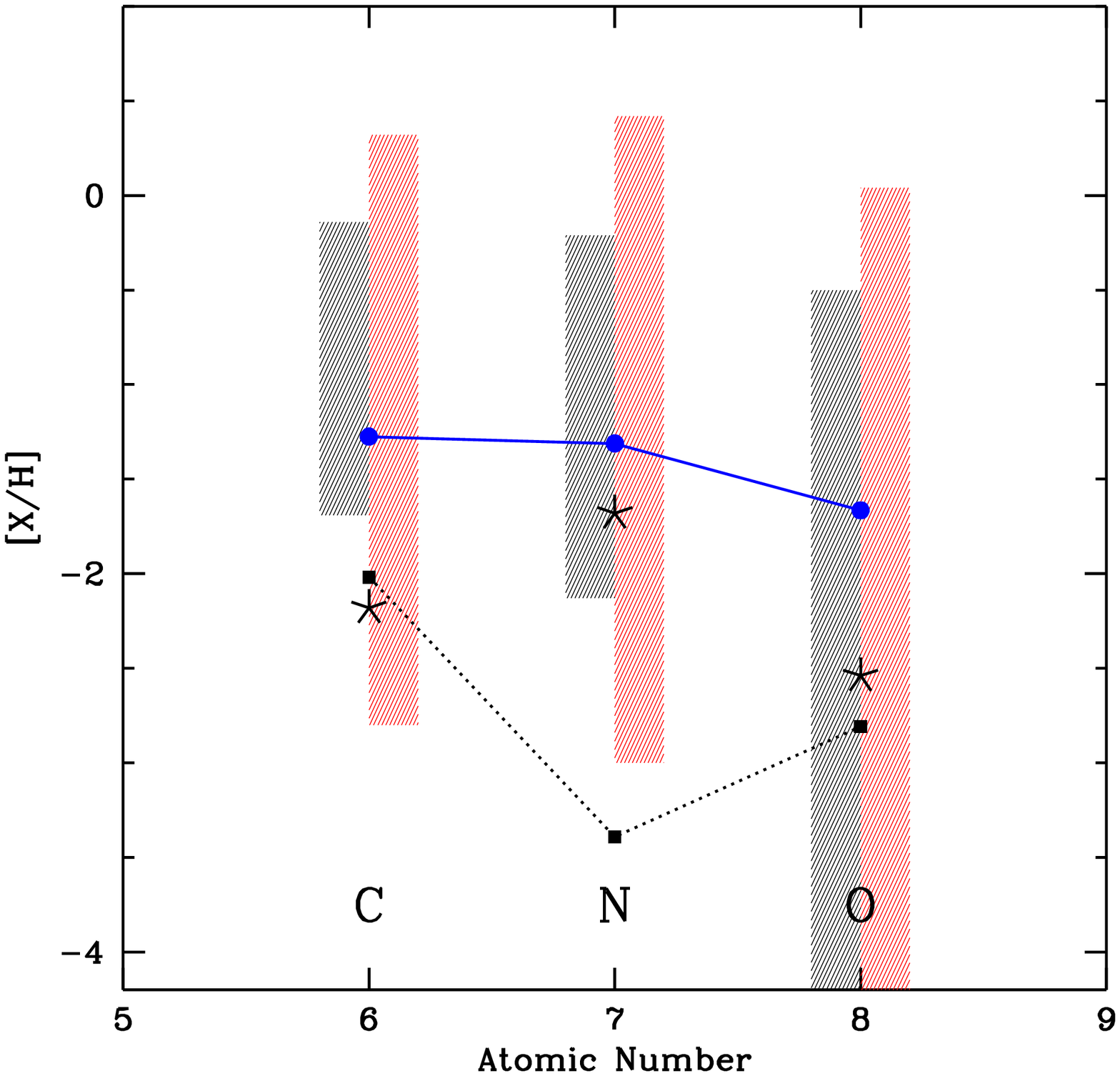}
\caption{{\it Left panel} : Composition of the mixture composed of wind material of  60 M$_\odot$ models at $Z=10^{-5}$
and of layers ejected by the supernova diluted with 10 times more interstellar material.  
The different curves correspond to the model WIND+SUPERNOVA in Table 2, various values of the mass-cut are
considered.
The continuous and short dashed lines curves correspond to the rotating model
     with $\upsilon_{\rm ini}=800$ km s$^{-1}$. The dotted lines and long dashed curves curves correspond
     to the non-rotating models \citep[models from][]{paperVIII2002}.  
 As the mass cut increases (downwards), the amount of carbon and oxygen ejected by the supernova decreases.  
 The stars show the observed values for the most iron poor star known today (Frebel et al. 2008).
     The vertical hatched zones show the range of observed values for CEMP-no stars with log $g$ superior to 3.8 (left grey zone, 5 stars in the sample)
     and with log $g$ inferior to 3.8 (right red zone, 20 stars in the sample). The values were taken from Tables 1 and 2 of \citet{Masseron2009}. Only upper limits are given for [O/H], this is why the columns extend down to the bottom of the figure.   
{\it Right panel :} The continuous line shows the
     composition obtained by mixing the outer envelope (mass above the CO core) at the E-AGB phase of our rotating 7 M$_\odot$ stellar model
     with 100 times more of interstellar medium. The dotted curve with squares shows the composition obtained
     in the same way (mass above the CO core and interstellar medium) using our analog non-rotating 7 M$_\odot$ stellar model.}
\label{comp}
\end{figure}

Let us call spinstars, stars whose evolution or nucleosynthesis is deeply affected by axial rotation. 
Such spinstars can contribute in two ways in shaping the chemical composition of the halo stars that
we observe today. First the ejecta of stars
of different masses, metallicities and initial rotation velocities can be mixed with ISM and provide
the raw material from which the ``chemically normal''  halo population stars are formed.

In order to compute the evolution with time or with [Fe/H] of such stars, detailed chemical evolution models
have to be used. The models by \citet{Chiappini2006} show that the N/O plateau observed at low $Z$
can be well reproduced when yields of rotating models ($V_{\rm ini}=800$ km s$^{-1}$) are used. 
The slower rotating models can produce a plateau for the
nitrogen to oxygen ratio at low metallicities but the level is two to three orders of magnitude below
the observed one. Non-rotating models would still be much lower and would not produce a plateau at low $Z$.
\citep[see Fig. 4][]{Chiappini2006}. Not only the N/O ratio is very well reproduced by these models but
also the C/O upturn mentionned in the introduction.  It would be extremely interesting to obtain information of the
$^{12}$C/$^{13}$C ratio in very metal poor stars, whose surface still reflects the abundances
of the cloud from which these stars were formed. \citet{Chiappini2008} show that the slow
and fast rotating models predict very different values for this ratio.

\begin{table}
\begin{center}
{\scriptsize
\caption{Chemical composition of the ejecta of various models in mass fraction. The number ratio $^{12}$C/$^{13}$C is also indicated.}
	\centering
		\begin{tabular}{cccccccccc}
		\hline
		\hline \\
M$_{\rm ini}$  & Z$_{\rm ini}$  & $\upsilon_{\rm ini}$       & M$_{\rm eje}$ & X$_{\rm H}$ & X$_{\rm He}$ 
& X$_{\rm C12}$ & $^{12}$C/ $^{13}$C      & X$_{\rm N14}$ & X$_{\rm O16}$    \\
	  \hline
\multispan{10}{\hfill WIND\hfill}	\\
                                  &                               &                                                         &                            &                       &                         &                     \\		   
	   85                 & 10$^{-8}$ & 800 &              65.2  & 0.27 & 0.55         & 1.0e-01            & 11.4             & 3.0e-02          & 5.0e-02      \\
	   60                 & 10$^{-8}$ & 800 &              36.2  & 0.40 & 0.59         & 1.3e-04            &  5.4              & 5.5e-03          & 1.7e-04      \\
	   40                 & 10$^{-8}$ & 700 &                4.2  & 0.68 & 0.32         & 1.9e-03            &  6.8              & 8.6e-04          & 5.8e-04      \\
                                  &                               &                                                         &                            &                       &                         &                     \\
	   60                 & 10$^{-5}$ & 800 &              23.1  & 0.47 & 0.53         & 1.4e-04            &  4.6              & 4.2e-04          & 1.2e-05      \\
                                  &                               &                                                         &                            &                       &                         &                     \\	   
\multispan{10}{\hfill WIND+SUPERNOVA (different mass cuts)\hfill}\\	
                                  &                               &                                                         &                            &                       &                         &                     \\
	   60                 & 10$^{-5}$ &      0 &               37  & 0.48 & 0.52         & 1.7e-07            &  3.65e+00              & 4.7e-06          & 2.1e-06      \\	   
	   60                 & 10$^{-5}$ &      0 &            41.8  & 0.42 & 0.57         & 6.4e-03            &  1.55e+05              & 4.3e-06          & 5.1e-03      \\
	   60                 & 10$^{-5}$ & 800 &               27  & 0.42 & 0.58         & 2.0e-05            &  4.70e+00              & 7.3e-04          & 1.7e-05      \\	   
	   60                 & 10$^{-5}$ &  800 &              29  & 0.39 & 0.61         & 3.4e-05            &  7.32e+00              & 8.5e-04          & 8.5e-05      \\
	   60                 & 10$^{-5}$ & 800 &               30  & 0.38 & 0.61         & 8.6e-04            &  1.90e+02              & 8.4e-04          & 6.9e-03      \\	   
	   60                 & 10$^{-5}$ &  800 &              32  & 0.37 & 0.60         & 2.4e-03            &  5.39e+02              & 8.1e-04          & 2.3e-02      \\
	   60                 & 10$^{-5}$ &  800 &           34.5  & 0.33 & 0.55         & 5.4e-03            &  1.37e+03              & 7.3e-04          & 1.0e-01      \\
                                  &                               &                                                         &                            &                       &                         &                     \\	   	   	   	   	   	   
\multispan{10}{\hfill E-AGB ENVELOPE\hfill}	\\	
                                  &                               &                                                         &                            &                       &                         &                     \\   
	    7                  & 10$^{-5}$ &      0 &                6.1  & 0.65 & 0.34         & 2.1e-03            & 4.30e+05      & 2.9e-06          & 2.0e-04      \\
	   7                   & 10$^{-5}$ & 800 &                5.7  & 0.63 & 0.35         & 9.6e-03            & 1.04e+02       & 3.2e-03          & 7.8e-03      \\		      
	\hline
	\end{tabular}
		\label{data}
		}
		\end{center}
\end{table}

Stars can also form from  interstellar material enriched by one or a very small number of nucleosynthetic events.
In that case one expects some important scatter from one star to another and also some strong differences with
respect to stars formed from the well mixed reservoir. The CEMP stars are likely formed in that way.

We show in Table \ref{data} the chemical composition of various components (winds, wind plus supernova ejecta, envelope of an early-AGB star)
 of material ejected by various stellar models. The composition indicated in that table would be the one of stars formed  from pure ejecta. 
Looking at the results of Table \ref{data}, the following trends can be deduced :
\begin{itemize}
\item Stars made of pure ejecta (or diluted with small amount of interstellar material) would be
He-rich. This is true whatever the source of the ejecta, rotating or non-rotating, wind material,
wind plus supernova material, or envelope of an early AGB star.
Such stars would be depleted in lithium. 
\item Low values for the $^{12}$C/$^{13}$C ratio indicates that the ejecta are rich in CNO-processed material.
This conclusion does not depend
on the degree of dilution with interstellar material in case we consider ejecta of spinstar models. Indeed.
in rotating models, both $^{12}$C and $^{13}$C are produced through primary channels in so large quantities 
that the dilution should be enormous for changing it.
\item We see that, when the contribution of the supernova increases, the $^{12}$C/$^{13}$C ratio becomes larger. Also the
abundances of carbon and oxygen increase. This is because
some layers, rich in He-burning products, are ejected. We conclude that stars presenting high [N/C] and [N/O] ratios cannot be mainly made
from material processed by He burning and thus from supernova ejecta (except if the supernova only eject the outermost layers). 
This is also independent of the degree of dilution with
interstellar material. It is also independent of the models considered (rotating or non rotating).
\item We see that rotating models are the only ones, among those presented in Table \ref{data}, showing strong enhancements of the three CNO
elements simultaneously. The reason for this is that, in these models, mixing of He-burning products with H-burning
products occurred in the star which has ejected the material. 
 \end{itemize}
Comparisons with observed values are made in Fig. \ref{comp}. Composition of pure ejecta mixed with some amount of interstellar material are plotted. 
An important point to mention in order to correctly interpret this figure is the following: for the rotating models, the positions of the different curves
in this diagram do not much depend on the initial metallicity. Here the quantities are plotted for models at $Z=10^{-5}$. Would we have plotted data
obtained from the rotating model at $Z=10^{-8}$, similar results would have been obtained. This is because all the three isotopes (and also $^{13}$C)
are produced by (quasi) metallicity independent channels. For the non-rotating models, we have a different situation. While $^{12}$C and $^{16}$O have
a strong primary component, $^{14}$N and $^{13}$C are only secondary. This means that, would we have plotted the results for a model
with $Z=10^{-8}$ for instance, the point corresponding to $^{14}$N would have shifted downwards by about 3 dex!  
Keeping this in mind, we see that
 that non-rotating massive stars
cannot fit the observed values of the Frebel star for instance ($Z \sim 10^{-6}$). 
In contrast, rotating models produce situations where the three elements
are enhanced. This is true for massive stars but also for intermediate mass stars.
We also see that with a dilution factor of 10, CEMP stars made up of winds and of a small amounts of supernova ejecta would
be He-rich. The $^{12}$C/$^{13}$C ratios would be as indicated in Table \ref{data} (the mass ejected is equal to 60 minus $M_{\rm cut}$).
Greater dilutions factors (about 10 times higher) seem to be required in case CEMP stars result from the mixture
of an E-AGB envelope with interstellar material. In that case, these stars would not be He-rich. They would also show high lithium abundances
at least equal to the Spite Li plateau in case the depletion process in those very metal poor stars occurred as
it occurred in the stars of the Spite plateau.
The $^{12}$C/$^{13}$C ratios would be equal to 105 and
2560 for the rotating and non-rotating case respectively.

\bibliographystyle{aa}
\bibliography{MyBiblio}

\begin{thebibliography}{32}
\expandafter\ifx\csname natexlab\endcsname\relax\def\natexlab#1{#1}\fi

\bibitem[{{Akerman} {et~al.}(2004){Akerman}, {Carigi}, {Nissen}, {Pettini}, \&
  {Asplund}}]{Akerman2004}
{Akerman}, C.~J., {Carigi}, L., {Nissen}, P.~E., {Pettini}, M., \& {Asplund},
  M. 2004, \aap, 414, 931

\bibitem[{{Andrievsky} {et~al.}(2009){Andrievsky}, {Spite}, {Korotin}, {Spite},
  {Fran{\c c}ois}, {Bonifacio}, {Cayrel}, \& {Hill}}]{Andrievsky2009}
{Andrievsky}, S.~M., {Spite}, M., {Korotin}, S.~A., {et~al.} 2009, \aap, 494,
  1083

\bibitem[{{Barkat} {et~al.}(1967){Barkat}, {Rakavy}, \& {Sack}}]{Barkat1967}
{Barkat}, Z., {Rakavy}, G., \& {Sack}, N. 1967, Physical Review Letters, 18,
  379

\bibitem[{{Beers} \& {Christlieb}(2005)}]{Beers2005}
{Beers}, T.~C. \& {Christlieb}, N. 2005, \araa, 43, 531

\bibitem[{{Bond} {et~al.}(1984){Bond}, {Arnett}, \& {Carr}}]{Bond1984}
{Bond}, J.~R., {Arnett}, W.~D., \& {Carr}, B.~J. 1984, \apj, 280, 825

\bibitem[{{Cayrel} {et~al.}(2004){Cayrel}, {Depagne}, {Spite}, {Hill}, {Spite},
  {Fran{\c c}ois}, {Plez}, {Beers}, {Primas}, {Andersen}, {Barbuy},
  {Bonifacio}, {Molaro}, \& {Nordstr{\"o}m}}]{Cayrel2004}
{Cayrel}, R., {Depagne}, E., {Spite}, M., {et~al.} 2004, \aap, 416, 1117

\bibitem[{{Cescutti}(2008)}]{Cescutti2008}
{Cescutti}, G. 2008, \aap, 481, 691

\bibitem[{{Chiappini} {et~al.}(2008){Chiappini}, {Ekstr{\"o}m}, {Meynet},
  {Hirschi}, {Maeder}, \& {Charbonnel}}]{Chiappini2008}
{Chiappini}, C., {Ekstr{\"o}m}, S., {Meynet}, G., {et~al.} 2008, \aap, 479, L9

\bibitem[{{Chiappini} {et~al.}(2006){Chiappini}, {Hirschi}, {Meynet},
  {Ekstr{\"o}m}, {Maeder}, \& {Matteucci}}]{Chiappini2006}
{Chiappini}, C., {Hirschi}, R., {Meynet}, G., {et~al.} 2006, \aap, 449, L27

\bibitem[{{Decressin} {et~al.}(2007{\natexlab{a}}){Decressin}, {Charbonnel}, \&
  {Meynet}}]{Decressin2007b}
{Decressin}, T., {Charbonnel}, C., \& {Meynet}, G. 2007{\natexlab{a}}, \aap,
  475, 859

\bibitem[{{Decressin} {et~al.}(2007{\natexlab{b}}){Decressin}, {Meynet},
  {Charbonnel}, {Prantzos}, \& {Ekstr{\"o}m}}]{Decressin2007a}
{Decressin}, T., {Meynet}, G., {Charbonnel}, C., {Prantzos}, N., \&
  {Ekstr{\"o}m}, S. 2007{\natexlab{b}}, \aap, 464, 1029

\bibitem[{{Ekstr{\"o}m} {et~al.}(2008{\natexlab{a}}){Ekstr{\"o}m}, {Meynet},
  {Chiappini}, {Hirschi}, \& {Maeder}}]{Ekstrom2008a}
{Ekstr{\"o}m}, S., {Meynet}, G., {Chiappini}, C., {Hirschi}, R., \& {Maeder},
  A. 2008{\natexlab{a}}, \aap, 489, 685

\bibitem[{{Ekstr{\"o}m} {et~al.}(2008{\natexlab{b}}){Ekstr{\"o}m}, {Meynet}, \&
  {Maeder}}]{Ekstrom2008d}
{Ekstr{\"o}m}, S., {Meynet}, G., \& {Maeder}, A. 2008{\natexlab{b}}, in IAU
  Symposium, Vol. 250, IAU Symposium, ed. {F.~Bresolin, P.~A.~Crowther, \&
  J.~Puls}, 209--216

\bibitem[{{Ekstr{\"o}m} {et~al.}(2008{\natexlab{c}}){Ekstr{\"o}m}, {Meynet},
  {Maeder}, \& {Barblan}}]{Ekstrom2008b}
{Ekstr{\"o}m}, S., {Meynet}, G., {Maeder}, A., \& {Barblan}, F.
  2008{\natexlab{c}}, \aap, 478, 467

\bibitem[{{Gratton} {et~al.}(2004){Gratton}, {Sneden}, \&
  {Carretta}}]{Gratton2004}
{Gratton}, R., {Sneden}, C., \& {Carretta}, E. 2004, \araa, 42, 385

\bibitem[{{Heger} \& {Woosley}(2005)}]{HegerWoosley2005}
{Heger}, A. \& {Woosley}, S. 2005, in IAU Symposium, Vol. 228, From Lithium to
  Uranium: Elemental Tracers of Early Cosmic Evolution, ed. {V.~Hill,
  P.~Fran{\c c}ois, \& F.~Primas}, 297--302

\bibitem[{{Hirschi}(2007)}]{Hirschi2007}
{Hirschi}, R. 2007, \aap, 461, 571

\bibitem[{{Honda} {et~al.}(2004){Honda}, {Aoki}, {Kajino}, {Ando}, {Beers},
  {Izumiura}, {Sadakane}, \& {Takada-Hidai}}]{Honda2004}
{Honda}, S., {Aoki}, W., {Kajino}, T., {et~al.} 2004, \apj, 607, 474

\bibitem[{{Ishimaru} {et~al.}(2004){Ishimaru}, {Wanajo}, {Aoki}, \&
  {Ryan}}]{Ishimaru2004}
{Ishimaru}, Y., {Wanajo}, S., {Aoki}, W., \& {Ryan}, S.~G. 2004, \apjl, 600,
  L47

\bibitem[{{Ishimaru} {et~al.}(2006){Ishimaru}, {Wanajo}, \&
  {Prantzos}}]{Ishimaru2006}
{Ishimaru}, Y., {Wanajo}, S., \& {Prantzos}, N. 2006, in International
  Symposium on Nuclear Astrophysics - Nuclei in the Cosmos

\bibitem[{{Karlsson} {et~al.}(2008){Karlsson}, {Johnson}, \&
  {Bromm}}]{Karlsson2008}
{Karlsson}, T., {Johnson}, J.~L., \& {Bromm}, V. 2008, \apj, 679, 6

\bibitem[{{Maeder}(2009)}]{Maeder2009}
{Maeder}, A. 2009, {Physics, Formation and Evolution of Rotating Stars}, ed.
  A.~Maeder

\bibitem[{{Maeder} \& {Meynet}(2001)}]{paperVII2001}
{Maeder}, A. \& {Meynet}, G. 2001, \aap, 373, 555

\bibitem[{{Maeder} {et~al.}(2009){Maeder}, {Meynet}, {Ekstr{\"o}m}, \&
  {Georgy}}]{Maeder2009a}
{Maeder}, A., {Meynet}, G., {Ekstr{\"o}m}, S., \& {Georgy}, C. 2009,
  Communications in Asteroseismology, 158, 72

\bibitem[{{Masseron} {et~al.}(2009){Masseron}, {Johnson}, {Plez}, {Van Eck},
  {Primas}, {Goriely}, \& {Jorissen}}]{Masseron2009}
{Masseron}, T., {Johnson}, J.~A., {Plez}, B., {et~al.} 2009, ArXiv e-prints

\bibitem[{{Meynet} {et~al.}(2006){Meynet}, {Ekstr{\"o}m}, \&
  {Maeder}}]{Meynet2006}
{Meynet}, G., {Ekstr{\"o}m}, S., \& {Maeder}, A. 2006, \aap, 447, 623

\bibitem[{{Meynet} \& {Maeder}(2002)}]{paperVIII2002}
{Meynet}, G. \& {Maeder}, A. 2002, \aap, 390, 561

\bibitem[{{Meynet} \& {Maeder}(2005)}]{paperXI2005}
{Meynet}, G. \& {Maeder}, A. 2005, \aap, 429, 581, paperXI

\bibitem[{{Pettini} {et~al.}(2008){Pettini}, {Zych}, {Steidel}, \&
  {Chaffee}}]{Pettini2008}
{Pettini}, M., {Zych}, B.~J., {Steidel}, C.~C., \& {Chaffee}, F.~H. 2008,
  \mnras, 385, 2011

\bibitem[{{Pignatari} {et~al.}(2008){Pignatari}, {Gallino}, {Meynet},
  {Hirschi}, {Herwig}, \& {Wiescher}}]{Pignatari2008}
{Pignatari}, M., {Gallino}, R., {Meynet}, G., {et~al.} 2008, \apjl, 687, L95

\bibitem[{{Ryan} {et~al.}(1996){Ryan}, {Norris}, \& {Beers}}]{Ryan1996}
{Ryan}, S.~G., {Norris}, J.~E., \& {Beers}, T.~C. 1996, \apj, 471, 254

\bibitem[{{Spite} {et~al.}(2005){Spite}, {Cayrel}, {Plez}, {Hill}, {Spite},
  {Depagne}, {Fran{\c c}ois}, {Bonifacio}, {Barbuy}, {Beers}, {Andersen},
  {Molaro}, {Nordstr{\"o}m}, \& {Primas}}]{Spite2005}
{Spite}, M., {Cayrel}, R., {Plez}, B., {et~al.} 2005, \aap, 430, 655

\end{thebibliography}

\end{document}